\input{aipcheck}

\documentclass[
    ,final            
  ]
  {aipproc}

\layoutstyle{6x9}

\newcommand{\hi}{H{\sc i}}

\newcommand{\kms}{km\,s$^{-1}$}

\begin{document}

\title{The EVLA: Prospects for \hi}

\keywords      {Radio Lines: ISM --- ISM: general --- Facilities: EVLA}

\author{J\"urgen Ott\footnote{J\"urgen Ott is a Jansky Fellow of the National Radio Astronomy Observatory; email: jott@nrao.edu}}{
  address={National Radio Astronomy Observatory, 520 Edgemont Road, Charlottesville, VA 22903, USA}, altaddress={California Institute of Technology, 1200 E California Blvd, Caltech Astronomy 105-24, Pasadena, CA 91125, USA}
}

\author{Rick Perley}{
  address={National Radio Astronomy Observatory, P.O. Box O, 1003 Lopezville Road, Socorro, NM 87801, USA }
}

\author{Michael Rupen}{
  address={National Radio Astronomy Observatory, P.O. Box O, 1003 Lopezville Road, Socorro, NM 87801, USA }
}

\author{EVLA team}{
  address={National Radio Astronomy Observatory, P.O. Box O, 1003 Lopezville Road, Socorro, NM 87801, USA }
}


\begin{abstract}
  To continue the unparalleled success of the Very Large Array (VLA)
  for radio astronomy, the facility is currently being converted to
  become the 'Expanded VLA' (EVLA). The EVLA will radically improve
  the VLA in order to cover the full 0.93-50\,GHz radio wavelength
  range without gaps, provide up to an order of magnitude better
  sensitivity, and to allow observations at much larger bandwidths and
  spectral resolution as currently possible. For observations of the
  21\,cm line of atomic neutral hydrogen (\hi), the EVLA offers
  thousands of \kms\ velocity coverage at sub-\kms\ resolution for
  targeted observations as well as an improved spectral baseline
  stability. In addition, every L--band (21\,cm) continuum or targeted
  \hi\ observation can be set-up to simultaneously observe a full
  $z=0-0.53$ \hi\ redshift survey at a velocity resolution of a few
  \kms. In turn, every \hi\ observation will also yield deep radio
  continuum images of the field. These synergies will deliver a wealth
  of data which opens up a wide 'discovery space' to study the details
  of galaxy evolution and cosmology.

\end{abstract}

\maketitle


\section{Introduction}

The Very Large Array (VLA) has played a key role in exploring the
radio Universe for almost three decades. To continue the unparalleled
success of this facility, the VLA is currently undergoing a drastic
rejuvenation process to become the 'Expanded VLA' or EVLA. The
conversion is very comprehensive and comprises additional and upgraded
receivers, new broad-band fiber optics, new online control systems,
new digital electronics, and a state-of-the-art correlator called
WIDAR (Wideband Interferometric Digital Architecture). The upgrades
will not only improve the continuum sensitivity by about an order of
magnitude largely due to a substantial increase of instantaneous
receiver and correlator bandwidth, but they will also provide extreme
spectral resolution and a wide coverage of the radio spectrum -- for
the first time it will be possible to observe at any chosen frequency
in the entire 1-50\,GHz radio window.

The EVLA specifications and its current status is described on the following webpage:\\
\verb=http://www.aoc.nrao.edu/evla=\\
(see also \cite{nap06}). With its unparalleled sensitivity the EVLA is
a true pathfinder for the Square Kilometer Array (SKA). Even the
EVLA's mapping speed matches that of special-purpose, ultra-wide field
but lower sensitivity and narrow band SKA pathfinders like ASKAP, ATA,
or MEERKAT in their early incarnations.

One of the fields of astronomy that the VLA pioneered is the
observation of atomic, neutral hydrogen (\hi) in our Milky Way, in
nearby galaxies and the intragroup medium, as well as at higher
redshifts. Those observations revealed the complex properties of the
neutral ISM, its importance on star formation and the physics of
the interface between a galaxy's disk, its halo and the intergalactic
medium. The abundance of \hi\ and its 21\,cm line properties also
proved to be indispensable tools to derive the complex dynamics of
galaxies in the form of, for example, density waves, rotation curves
(and thus the dark matter distribution), tidal interactions, and
mergers of galaxies. In this article we like to summarize what EVLA
offers for future \hi\ observations.

\section{EVLA Receivers for \hi\ Observations}

The \hi\ hyperfine line rest frequency of $\sim 1.420$\,GHz falls into
the radio L--band. The VLA L--band receivers cover a frequency range
of 1.25--1.8\,GHz. The EVLA receivers will widen this to
0.93--2.1\,GHz. At the lower end this will increase the \hi\ redshift
coverage from a current upper limit of $z\sim 0.14$ to $z\sim
0.53$. This is an improvement of almost $\Delta z\sim 0.4$ or an
additional $\sim 3.5$\,Gyr of look-back time (in a WMAP $\Lambda$CDM
cosmology). The system temperature over telescope efficiency T$_{\rm
  sys}/\epsilon$ of the EVLA is expected to decrease from $\sim 75$\,K
to about 60\,K or less (T$_{\rm sys}$ of the EVLA is designed to hover
around $\sim 26$\,K), saving about 30\% of integration time to reach
the same sensitivity.

The old VLA L--band feedhorn design featured a microwave lens in the
optical path. With the redesigned EVLA receivers, this lens is not
required anymore. Thus, ground radiation scattered by the lens (and
other structural dish elements) into the feedhorn is largely reduced
and the EVLA L--band system temperature improves substantially over
the VLA at lower elevations; at an elevation of $\sim 20^{\circ}$ the
EVLA system temperature is about half that of the VLA which
corresponds to a four fold increase in sensitivity.

As part of the EVLA conversion, the L--band receivers will be equipped
with new orthomode transducers. Until they become available for all
antennas in 2012, an interim L--band system is currently being
installed (with a slightly lower sensitivity and a minimum frequency
of 1\,GHz, corresponding to $z\sim0.42$). For more information on the
performance of the EVLA L--band system upgrade and performance, we
would like to refer to the EVLA webpages and also to Emmanuel
Momjian's contribution in this volume.

At lower frequencies the EVLA offers a P--band receiver which covers
frequencies of 300--340\,MHz equaling an \hi\ $z\sim 3.2-3.7$
redshift range. This band remains unchanged in the VLA to EVLA
conversion. The system is not sensitive enough to observe typical, gas
rich galaxies at the available redshift in \hi\ emission. However,
searches for \hi\ absorption in P--band have been conducted in the
past (e.g.\cite{tar94}) and are still an option for the EVLA.

\section{The WIDAR correlator}

Since the installation of the VLA correlator, Moore's law pushed
processing speeds of computers by $\sim 5$ orders of magnitude. Taking
advantage of this, a new correlator, WIDAR, will be commissioned in
2008/2009. However, one should keep in mind that the current VLA correlator
was designed to be very suitable for \hi\ observations toward nearby
galaxies. The bandwidth and resolution almost perfectly matches what
is needed for such observations (bandwidth of $\sim 200$\,\kms\ at a
resolution of $\sim 5$\,\kms). But the VLA correlator design severely
limits the amount of 'discovery space', e.g., very wide and shallow
lines (for an example, see \cite{mor07}) would not be discovered with
the VLA without prior knowledge of these features. The narrow
bandwidth typically used for galaxies of $\sim 1.5$\,MHz has only few,
supposedly line--free channels at the band edges. Any wide lines
extending across these channels would not be discovered because they
would be removed in the process of continuum subtraction. Another case
of limited discovery space is that other \hi\ sources in the field,
e.g., companion galaxies, remain undetected with the VLA if they are
at a slightly different velocity than the main target. At the other
extreme, very narrow line features, e.g., caused by \hi\
self--absorption are smeared out and would be missed in a typical
extragalactic \hi\ setup with its velocity resolution of a few
\kms. To open up new discovery space, wide bands at high spectral
resolution are desired. Such capabilities are provided by the new
WIDAR correlator (for a description of the technology, see
\cite{car00}). WIDAR will have a spectral resolution of down to Hz
ranges and a bandwidth of up to 8\,GHz. The full bandwidth is split up
into four 2\,GHz baseband pairs and in each baseband pair up to 16
independent sub-band pairs can be selected with bandwidths between
31.25\,kHz and 128\,MHz. The full 8\,GHz baseband pairs have a {\it
  minimum} of 16384 channels which will always be
available. Recirculation trades bandwidth for more channels and the
maximum number of channels are of order 4 million. Such a flexible
design will cover virtually any need for setups to observe \hi\ in
single targets with thousands of \kms\ bandwidth and sub-\kms\
velocity resolution.

But the EVLA will be able to do more. The L--band receivers and WIDAR
will cover the entire \hi\ redshift range of $z=0-0.53$ at a
resolution of 3.2\,\kms\ when observed with two polarization products,
and at 6.4\,\kms\ when observed at full stokes. Other configurations
will be able to, e.g., stack multiple radio recombination lines in
order to improve the signal--to--noise of Zeeman splitting
experiments in a single observation.

\section{Piggy-backing}

The velocity resolution of a few \kms\ over the full 0.93-2.1\,GHz
L--band range enables new, unique synergies with other
observations. Every EVLA L--band continuum observation will also be an
\hi\ redshift survey and vice versa. Also, virtually every targeted
\hi\ observation leaves enough computing power in WIDAR to once more
perform a simultaneous \hi\ survey over the entire $z=0-0.53$ range at
good velocity resolution. As if this would not be enough, the minimum
data dumping time for 1\,GHz bandwidth L--band data is $\sim 100$\,ms
which allows monitoring of and searches for transient sources in the
field while simultaneously observing \hi\ or radio continuum projects
(but note that current data output limitations imposed by archiving
are $\sim 25$\,MB\,s$^{-1}$, equaling to dumping times of or $\sim
20$\,s, when the maximum number of channels and baselines is read
out). These are exciting new opportunities that will pave the way
toward SKA \hi\ surveys. For example, without spending any dedicated
survey time, the $z=0-0.53$ redshift volume around a VLA standard
calibrator will accumulate hundreds of hours prior to the
commissioning of the SKA. This will provide very deep \hi\ and radio
continuum images essentially for free. The up to 8 bit quantization of
WIDAR also delivers improved high dynamic range imaging capabilities
which reduce current sensitivity limitations due to the inevitable
presence of strong sources in any field. The new digital transmission
system also removes the infamous ``3 MHz ripple'' and related spectral
baseline instabilities. This reduces the systematic uncertainties of
deep \hi\ observations dramatically.

\section{Summary}

Over the VLA, the EVLA will improve \hi\ observations in terms of a
wider L--band frequency range (down to $\sim 930$\,MHz), a better
$T_{\rm sys}/\epsilon$ sensitivity (in particular at off-zenith
elevations), much improved spectral baseline stability, and, most
importantly, spectral bandwidth and resolution. The L--band receiver
improvement guarantee that the EVLA will still remain the most
sensitive interferometer for \hi\ in the world for at least a decade,
until SKA pathfinders will be expanded far beyond the currently
planned prototypes. The new WIDAR correlator is flexible enough to
allow observations of virtually any galaxy at sub--\kms\ resolution
with thousands of \kms\ bandwidth. At the same time, every L--band
continuum or targeted \hi\ observation will also yield a blind
$z=0-0.53$ redshift \hi\ survey at a velocity resolution of a few
\kms, and vice versa. It is clear that these new possibilities have
their price in a very large data rate and that data reduction will
challenge today's computing capabilities. The opportunities, however,
are tremendous and it is up to the community to develop new strategies
on how to take advantage of the wealth of EVLA data in order to answer
the open questions of galaxy evolution and cosmology.

%
%
\begin{theacknowledgments}
The National Radio Astronomy Observatory is a facility of the National Science Foundation operated under cooperative agreement by Associated Universities, Inc.
\end{theacknowledgments}
%
%

\bibliographystyle{aipproc}   

\bibliography{sample}

\IfFileExists{\jobname.bbl}{}
 {\typeout{}
  \typeout{******************************************}
  \typeout{** Please run "bibtex \jobname" to optain}
  \typeout{** the bibliography and then re-run LaTeX}
  \typeout{** twice to fix the references!}
  \typeout{******************************************}
  \typeout{}
 }




\end{document}